\documentclass[prl,twocolumn,showpacs,superscriptaddress]{revtex4}

\usepackage{graphicx}
\usepackage{epsf,latexsym}
\usepackage{times}

\newcommand{\be}{\begin{equation}}
\newcommand{\ee}{\end{equation}}
\newcommand{\ba}{\begin{eqnarray}}
\newcommand{\ea}{\end{eqnarray}}
\newcommand{\up}{\uparrow}
\newcommand{\dw}{\downarrow}
\newcommand{\ud}{{\uparrow\downarrow}}

\newcommand{\si}{\sigma}
\newcommand{\sib}{{\bar{\sigma}}}
\newcommand{\sip}{{\sigma'}}
\newcommand{\vv}[1]{{\bf #1}}
\newcommand{\mchi}{\underline{\underline{\bf\chi}}\,}
\newcommand{\alda}{^{\scriptscriptstyle \rm ALDA}}
\newcommand{\halda}{^{\scriptscriptstyle \rm H+ALDA}}
\newcommand{\VE}{^{\scriptscriptstyle \rm VE}}
\newcommand{\scd}{^{\scriptscriptstyle \rm SCD}}

\newcommand{\sinf}{\raisebox{-.7ex}{$\stackrel{<}{\sim}$}}

\begin{document}

\title{Dissipation through spin Coulomb drag in electronic spin dynamics}

\author{I. D'Amico}
\email{ida500@york.ac.uk}
\affiliation{Department of Physics, University of York, York YO10 5DD, United Kingdom}

\author{C. A. Ullrich}

\affiliation{Department of Physics and Astronomy, University of Missouri-Columbia,
Columbia, Missouri, 65211}

\date{\today }

\begin{abstract}
Spin Coulomb drag (SCD) constitutes an intrinsic source of dissipation for spin currents in metals and semiconductors.
We discuss the power loss due to SCD in potential spintronics devices and analyze
in detail the associated damping of collective spin-density excitations. It is found that SCD contributes
substantially to the linewidth of intersubband spin plasmons in parabolic quantum wells, which suggests the
possibility of a purely optical quantitative measurement of the SCD effect by means of inelastic light scattering.
\end{abstract}






\maketitle

Spintronics applications are receiving increasing attention in the hope of revolutionizing traditional technology
by a powerful exploitation of the spin -- as well as the charge -- degrees of freedom. An intense research
effort is underway to improve our understanding of spin dynamics, especially related to nanocircuits and their components,
such as quantum wells and wires. In this context the theory of spin Coulomb drag (SCD) was recently
developed \cite{SCD2000,SCD2001,SCD2002,SCDfle,SCD2003}.
This theory analyzes the role of Coulomb interactions between different spin populations in
spin-polarized transport. Coulomb interactions transfer momentum between different spin populations, so that
the total momentum of each spin population is not preserved. This provides an {\it intrinsic} source of friction
for spin currents, a measure of which is given by the spin-transresistivity \cite{SCD2000}.
SCD is generally small in metals, due to a typical Fermi temperature of the order of $10^5$ K, but  can
become substantial in semiconductors, where the spin-transresistivity can be larger than the Drude
resistivity \cite{SCD2002,SCD2003}. As the quest for defect-free  materials with longer and longer
spin-decoherence times is continuing, spurred by  practical requirements in spintronics as well as in quantum
computation devices, the SCD is bound to become one of the most serious issues in spin polarized transport, since,
due to its intrinsic nature, it cannot be avoided even in the purest material.
In fact, the recent experimental observation of SCD by Weber {\em et al.} \cite{Ore2005}
shows that the effect dominates spin diffusion currents over a broad range of parameters,
in agreement with theoretical predictions \cite{SCD2001,SCD2002,SCD2003}.

In this letter we discuss a critical issue for potential spintronics devices, namely
the power loss in spin transport and dynamics due to SCD. We shall analyze in detail its effect on
optical spin excitations, and propose an experiment to measure the intrinsic SCD linewidth enhancement
of spin plasmons in parabolic semiconductor quantum wells. While up to now SCD
has been considered only in relation to  spin transport, the proposed experiment
would provide an alternative way of measuring this subtle effect, and thus
establish unequivocally the influence of SCD on optical excitations.

Let us consider a system composed of spin-up and spin-down electron populations, as for example
the electrons in the conduction band of a doped semiconductor structure. We are assuming
spin-flip times long enough so that spin populations are well defined on the relevant time scales.
This assumption -- at the very core of spintronics --  has been proved reasonable, with experimentally
measured spin-decoherence times of the order of microseconds \cite{Awschanature}.
Previous papers on SCD have mainly analyzed the dependence of the spin-transresistivity over
temperature \cite{SCD2001,SCD2002,SCDfle,SCD2003}; this letter will focus on its frequency dependence
\cite{SCD2000}, which is important both for AC spintronics
applications and spin-resolved optical experiments.

In the linear response regime and for weak Coulomb coupling
one can write a phenomenological equation of motion for the spin $\sigma$ population \cite{SCD2000}.
The SCD force is defined as the Coulomb force (per unit volume) exerted by spin $\bar{\sigma}
(= - \sigma)$ electrons, moving with velocity ${\bf v}_{\bar{\sigma}}$, on spin $\sigma$ electrons,
moving with velocity ${\bf v}_{\sigma}$:
\begin{equation}
{\bf F}_{ \sigma\bar{\sigma}}(\vv{r};\omega)=
- \gamma(\omega) m  \frac{n_\sigma n_{\bar\sigma}}{n}\left[{\bf
v}_\sigma(\vv{r})-{\bf v}_{\bar{\sigma}}(\vv{r})\right], \label{gamma1}
\end{equation}
where the number density, $n_\sigma$, of $\sigma$-spin electrons of effective mass $m$,
and the total density, $n=n_\uparrow + n_\downarrow$, are those of a homogeneous reference system.
The drag coefficient $\gamma$ appearing in Eq.~(\ref{gamma1}) is directly proportional to the real part of the
spin-transresistivity $\rho_\ud$\cite{SCD2000}:
\begin{equation}
\gamma (\omega,T)= - \frac{ne^2}{m } \: \Re\rho_{\uparrow
\downarrow}(\omega,T;n_\up,n_\dw) \:,
\label{gamma}
\end{equation}
where $T$ is the electronic temperature. $\Re\rho_\ud$ has a negative value and $\rho_\ud$ can be defined by the
relation $\left.{\bf E}_\up \right|_{{\bf j}_{\up}=0}=-e {\bf j}_{\dw} \rho_\ud$,
with ${\bf j}_{\sigma}$ the number current density of the $\sigma$ spin population, ${\bf E}_\up$ the effective
electric field  which couples to the $\up$-spin population and includes the gradient of the local chemical
potential, and $e$ the absolute value of the electronic charge.

As noted above, SCD provides an intrinsic decay mechanism for spin-polarized currents, and is thus a
source for power loss in a spintronics circuit or device.
From the general definition of power and using Eq.~(\ref{gamma1}),
the SCD power loss density per unit time for the $\si$-spin population is given by
\ba
P_{\si}({\bf r};\omega,n_\up, n_\dw) &=&  {\bf F}_{ \sigma\bar{\sigma}}
({\bf r})\cdot{\bf v}_\si({\bf r})\label{power_loss_gen} \\
& = & e^2\left[\frac{n_\sib }{ n_\si}\left|{\bf j}_\si({\bf r})\right|^2-{\bf j}_\sib({\bf r})
\cdot{\bf j}_\si({\bf r})\right] \nonumber\\
&&\times \Re\rho_{\uparrow \downarrow}(\omega,T;n_\up, n_\dw) \,. \label{P_om_su_V}
\ea
Notice that  $P_{\si}$ can change sign depending on the {\it relative strength and direction} of the spin-resolved
current densities, a positive sign implying that the $\si$ spin population is being dragged along by the faster $\sib$
spin population. In particular, for a system with spin populations drifting at the same average velocity,
$P_\si({\bf r};\omega)=0$. The total power loss per unit time in a system with a slowly varying density
can be calculated as
\be  \bar{P}_\si(\omega)
=  \int_V d^3r \left[P_{\si}({\bf r};\omega,n_\up({\bf r}), n_\dw({\bf r}))\right]. \label{P_om}
\ee

\begin{figure}
\unitlength1cm
\begin{picture}(5.0,5.25)
\put(-4.25,-2.9){\makebox(5.0,5.25){\includegraphics{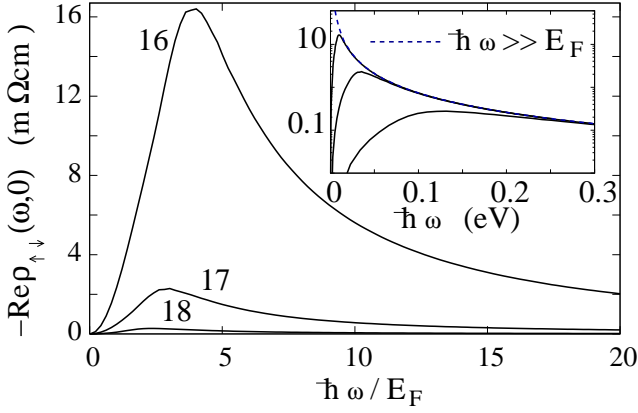}}}
\end{picture}
\caption{
Spin-transresistivity $|\Re\rho_\ud|$ vs rescaled frequency
$\hbar\omega/E_F$ for $n=10^{x}$ cm$^{-3}$, $x=16,17,18$ as indicated,
and GaAs parameters ($m=0.067 m_e$, $\epsilon=12$). Inset: $|\Re\rho_\ud|$ in $\rm m\Omega \,cm$ vs
$\hbar\omega$ in eV for the same densities. Dashed line: high-frequency limit, Eq. (\ref{asym}).
}
\label{fig1}
\end{figure}

Fig. \ref{fig1} shows the transresistivity $\Re\rho_\ud(\omega;n_\up, n_\dw)$ as a
function of frequency, calculated for GaAs at $T=0$, using a generalized random phase approximation \cite{SCD2000}.
We see that $\Re\rho_\ud$ has a maximum when
$E_{F\si}(n_\si(z))$ is of order $ \hbar\omega$ ($E_{F\si}$ is the
$\si$-spin Fermi energy). This maximum roughly scales as \cite{SCD2002}
$(ha^*/e^2)/n^s\approx 140 \, \mu \Omega \, {\rm cm}\cdot \epsilon m_e/(mn^s) $
with $s\sinf 1$: it is then reasonable to expect a sizable damping effect due to SCD.
We notice also that for very low densities, i.e. $E_F\ll\hbar\omega$,
\be \label{asym}
\Re\rho_\ud(\omega,T=0;n_\up,
n_\dw)\sim-{\hbar a^*\over e^2}\left({2Ry^*\over\hbar\omega}\right)^{3/2}{4\pi\over3},\label{rho_high_om}
\ee
independent of the carrier density (see Fig. \ref{fig1} inset)\cite{qian}.

Due to problems with electrical injection \cite{Schmidt} and the necessity of driving spin dynamics on
sub-picosecond time-scales \cite{QC}, large attention has been focused on optical
spin injection \cite{Awschanature} and optically controlled spin dynamics \cite{Awschaoptiflip};
in the following, we will explore how the SCD affects the lifetime and dynamics of
spin-dependent optical excitations.

The excitation spectrum of a system can be  calculated in principle exactly with time-dependent
density-functional theory (TDDFT) \cite{Petersilka}. In TDDFT, the properties of an interacting
time-dependent many-body system are described through a  non-interacting time-dependent
system (the so-called Kohn-Sham system) characterized by an exchange-correlation (xc) potential.
The  xc potential is a functional of the current \cite{VK}, and needs to be approximated in practice.
An approximation which is cast in the language of hydrodynamics, including dissipation effects, was proposed in Ref.
\cite{VE}: nonadiabatic xc effects manifest themselves as viscoelastic
stresses in the electron liquid, which are proportional to the velocity gradient. The corresponding expression for
spin-dependent systems is discussed in \cite{qian}, the main difference being the appearance of a term --
in addition to the viscoelastic tensor -- representing the damping of the spin currents due to the SCD effect.

Our derivation of the excitation energies for a spin-depen\-dent system closely follows the spin-independent
case, see Ref. \cite{UB} for details. Starting point is the TDDFT current response equation,
\begin{equation} \label{j_resp}
{\bf j}_\sigma({\bf r},\omega) = \frac{e}{c}\int \!d^3 r' \: \mchi_\sigma({\bf r},{\bf r}',\omega)
{\bf a}_\sigma({\bf r}',\omega) \;.
\end{equation}
Here, $\mchi_{\si}({\bf r},{\bf r}',\omega)$ is the Kohn-Sham current-current
response tensor, which is diagonal in the spin-channel.
The effective vector potential is defined as $\vv{a}_\si = \vv{a}_\si^{\rm ext} + \vv{a}_\si^{\rm H} + \vv{a}_\si^{\rm xc}$,
where $\vv{a}_\si^{\rm ext}$ is an external perturbation, and the Hartree and xc vector potentials are given by
\begin{equation}
\frac{e}{c}\: a_{\nu\sigma}^{\rm H}({\bf r},\omega) = \frac{\nabla_\nu}{(i\omega)^2}
\int \! d^3r' \: \frac{\nabla' \cdot {\bf j}({\bf r}',\omega)}{|{\bf r} - {\bf r}'|} \:,
\end{equation}
\begin{eqnarray} \label{vectorpot}
\frac{e}{c}\: a_{\nu\sigma}^{\rm xc}({\bf r},\omega)
&=& \!\!
\sum_{\sigma'}\frac{\nabla_\nu}{(i\omega)^2}\!  \int \! d^3 r' \,\nabla ' \cdot {\bf j}_{\sigma'}({\bf r}',\omega)\:
f_{\rm xc,\sigma \sigma'}\alda({\bf r},{\bf r}') \nonumber\\
&-&\frac{1}{i\omega n_\sigma({\bf r})} \sum_{\kappa\sip} \nabla_\kappa \sigma^{\rm xc}_{\nu\kappa,\sigma\sip}({\bf r},\omega)
\nonumber\\
&-& \frac{e^2}{\omega}\: n_\si({\bf r}) n_\sib({\bf r}) \rho_\ud(\omega;n_\si({\bf r}) n_\sib({\bf r}))
\nonumber\\
&&\times
\sum_{\sigma'} \frac{\sigma \sigma'}{n_\sigma({\bf r}) n_{\sigma'}({\bf r}) } \: j_{\nu \sigma'}({\bf r},\omega) \:,
\end{eqnarray}
where $\nu,\kappa$ are Cartesian indices.  In Eq. (\ref{vectorpot}),
\begin{equation}
f_{\rm xc,\sigma \sigma'}\alda({\bf r},{\bf r}') = \delta({\bf r} - {\bf r}')
\left. \frac{d^2e_{\rm xc}^{h}(\bar{n}_\uparrow,\bar{n}_\downarrow)}
{d\bar{n}_\sigma d\bar{n}_{\sigma'}} \right|_{\displaystyle \bar{n}_{\uparrow,\downarrow}=n_{0\uparrow,\downarrow}({\bf r})}
\end{equation}
is the frequency-independent xc kernel associated with the adiabatic local-density approximation (ALDA), where
$e_{\rm xc}^h$ is the xc energy density of a homogeneous electron gas, and $n_{0\sigma}$ the ground-state spin
density of the system. The other terms in Eq. (\ref{vectorpot}) represent non-adiabatic xc contributions, which
bring in the dissipation.  In the second term, $\sigma^{\rm xc}_{\nu\kappa,\sigma\sip}$ is the spin-resolved
viscoelastic stress tensor of the  electron liquid \cite{qian}. The key quantity in the last term of
Eq.~(\ref{vectorpot}) is $\rho_\ud$.

We now consider a specific excitation $p\sigma\to q\sigma$ between the Kohn-Sham levels
$\psi_{p\sigma}$ and $\psi_{q\sigma}$, and assume the ground state to be spin unpolarized.
To derive the TDDFT correction to the bare Kohn-Sham excitation
energy $\hbar \omega_{pq\sigma}$, we apply the so-called small-matrix approximation \cite{UB,appel}.
The result is, to lowest order in the non-adiabatic corrections,
\ba
\hbar^2\omega^2_{\pm\sigma} &=& \hbar^2\omega_{pq\sigma}^2+ 2\hbar \omega_{pq\sigma}\left[( S\halda_{\si\si}
\pm S\halda_{\sib\si}) \right.\nonumber\\
&& \left. {} + ( S\VE_{\si\si} \pm S\VE_{\sib\si})+ ( S\scd_{\si\si} \pm S\scd_{\sib\si})\right], \label{omsqgen}
\ea
where the $+/-$ sign refers to charge- or spin-density excitations (CDE/SDE) respectively. The terms in square brackets,
$S\halda_{\si\sip}$, $S\VE_{\si\sip}$ and $S\scd_{\si\sip}$, are the dynamical many-body corrections to the bare
transition energy $\hbar \omega_{pq\sigma}$ between the single particle  levels $p\sigma$ and $q\sigma$.
The Hartree+ALDA shift is given by
\begin{eqnarray}
S\halda_{\si\si'} &=& \int\!d^3r \! \int \!\! d^3 r' \psi_{p\sigma}({\bf r}) \psi_{q\sigma}({\bf r})
\psi_{p\sigma'}({\bf r}') \psi_{q\sigma'}({\bf r}')\nonumber\\
&&\times \left[\frac{1}{|{\bf r}-{\bf r}'|}
+ f_{\rm xc,\sigma \sigma'}\alda({\bf r},{\bf r}') \right],
\end{eqnarray}
which causes no dissipation, $f_{\rm xc,\sigma \sigma'}\alda$ being frequency independent
and real. The viscoelastic shift is given by
\begin{equation} \label{VE}
S\VE_{\si\si'} = \frac{i \omega}{\omega_{pq\sigma}^2 } \sum_{\nu\kappa}\int\! d^3 r \:
\sigma^{{\rm xc},pq}_{\kappa\nu,\si\si'}({\bf r},\omega)
\nabla_\kappa  \left[\frac{j_{pq\sigma,\nu}({\bf r})}{n_\si({\bf r})} \right],
\end{equation}
where $\sigma^{{\rm xc},pq}_{\kappa\nu,\si\sip}$ is the xc stress tensor \cite{VE,qian,UB} with the
exact current ${\bf j}_{\si,\nu}$ replaced by  ${\bf j}_{pq\si}({\bf r})\equiv \langle
\psi_{p\si}|{\hat {\bf j}_{\si}}|\psi_{q\si}\rangle$, with ${\hat {\bf j}_{\si}}$ the paramagnetic particle current density operator.
Eq.~(\ref{VE}) can be viewed as the average rate of energy dissipation per unit time in a viscous fluid \cite{UB,landau},
where $\sigma^{{\rm xc},pq}_{\kappa\nu,\si\sip}$ is the  viscoelastic stress tensor of the fluid, and
$\nabla_\kappa [{j_{pq\sigma,\nu}/ n_\si}]$ the velocity gradient. In contrast to the familiar expression from
classical fluid dynamics \cite{landau}, $S\VE$ has both real and imaginary part.

The SCD shift is a central result of this letter:
\ba
\label{SCD}
S\scd_{\si\si} \pm S\scd_{\sib\si}&=& \frac{i e^2\omega}{\omega_{pq\si}^2 }\int\! d^3 r \: \rho_{\uparrow
\downarrow}(\omega;n_\up({\bf r}), n_\dw({\bf r}))\nonumber \\
&\times& \!\!\! \left[\frac{n_\sib({\bf r})}{ n_\si({\bf r})}
\left|{\bf j}_{pq\si}({\bf r})\right|^2\mp{\bf j}_{pq\sib}({\bf r})
\cdot{\bf j}_{pq\si}({\bf r})\right] \!. \quad
\ea
As we will show in an example below, under certain circumstances this new contribution to the broadening
of an excitation can actually dominate the damping process.

By comparison with Eqs.~(\ref{P_om_su_V}) and (\ref{P_om}),  we immediately recognize
the structure of the power loss typical of the Coulomb drag force \cite{note2}.
Like the viscoelastic term (\ref{VE}), the SCD term (\ref{SCD}) contains both a real and an imaginary part.
Notice that, if the external driving force couples in a different way to the two
spin components, such that the average spin velocities are different, the SCD term contributes to the charge channel
too. In this particular case the two spin-populations may be considered
distinguishable, characterized by a spin-dependent frequency $\omega_\si$ both in the charge and
in the spin channel. This implies that  the Coulomb drag force exerted by one population onto the other
can be regarded as an external force.

\begin{figure}
\unitlength1cm
\begin{picture}(5.0,6.5)
\put(-3.75,-3.5){\makebox(5.0,6.5){\includegraphics{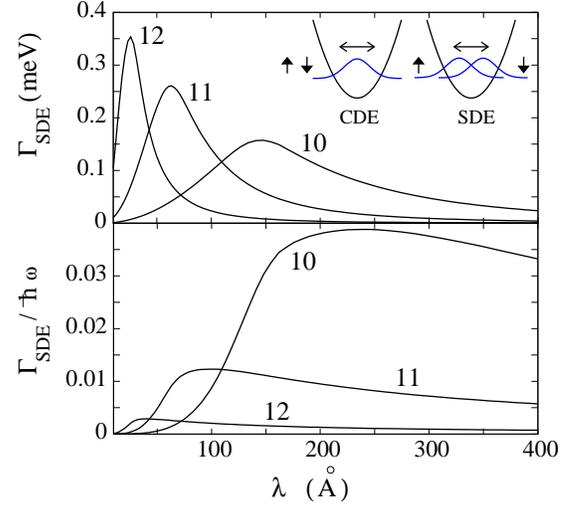}}}
\end{picture}
\caption{Upper panel:Spin-plasmon linewidth $\Gamma_{\rm SDE}^{\rm SCD}$ for a parabolic quantum well
versus curvature parameter $\lambda$,
for $N_s=10^{10}$, $10^{11}$ and $10^{12}$ cm$^{-2}$ and GaAs parameters. The inset illustrates the collective
motion of the two spin populations (CDE: in phase, SDE: out of phase).
Lower panel: rescaled linewidth $\Gamma_{\rm SDE}^{\rm SCD}/\hbar\omega$ versus $\lambda$ for the same system and parameters as upper panel.
}
\label{fig2}
\end{figure}

This concept can be clarified by considering the charge and spin plasmons
in a quantum well \cite{marmorkos,ullrichflatte1,ullrichflatte2}. The inset to Fig. \ref{fig2} illustrates the two types of
density oscillations for a parabolic well, in which the $n_\uparrow$ and $n_\downarrow$ components move back
and forth in phase (CDE) or with opposite phase (SDE). In the case of the SDE, the average net momentum
transferred by Coulomb interactions from the $\sib$ to the $\si$-spin population will be directed opposite
to the $\si$-spin direction of motion, so that the SCD effect damps the motion of {\it both} spin populations.
For the charge plasmon the effect can become more subtle: since the average spin velocities are in the same direction,
the net result of Coulomb interactions between the two spin populations will be to transfer momentum from the
"hotter" to the "colder" population, until equilibrium is reached. In this case the SCD effect would {\it not} damp
the motion of both spin populations, but pump momentum from the faster to the slower.

We now proceed to estimate the size of the SCD effect for optical excitations in a parabolic quantum well.
According to the Harmonic Potential Theorem \cite{dobson}, the intrinsic linewidth of a CDE
in a parabolic confining potential is strictly zero. The TDDFT linear response equation (\ref{j_resp})
satisfies this requirement: CDE's in a parabolic well have a uniform velocity
profile, so that the viscoelastic stress tensor vanishes. Likewise, in  expression
(\ref{VE}) for $S_{\sigma\sigma'}\VE$, $\nabla_\kappa[j_{pq\sigma,\nu}/n_\sigma]$ is very small.
The viscoelastic contributions to SDE's are thus a higher-order correction
compared to the SCD contributions, which give the dominant correction to the excitation frequency beyond ALDA.
The intrinsic SDE linewidth for a parabolic quantum well therefore becomes
$\Gamma_{\rm SDE}\approx \Gamma_{\rm SDE}^{\rm SCD}$, where
\ba
\Gamma_{\rm SDE}^{\rm SCD}(\omega) &=& \Im \left[S\scd_{\si\si} - S\scd_{\sib\si}\right]\nonumber\\
&=&\frac{ e^2N_s\omega}{2\omega_{pq\si}^2 }\int\!
dz \: \Re\rho_\ud(\omega;n_\up(z), n_\dw(z))\nonumber \\
&\times& \!\!\! \left[\frac{n_\sib(z)}{n_\si(z)}\left|{\bf j}_{pq\si}(z)\right|^2
+{\bf j}_{pq\sib}(z)
\cdot{\bf j}_{pq\si}(z)\right] \! , \quad
\label{gammaqw}
\ea
with $N_s$ the two-dimensional electronic sheet density.

Numerical results for $\Gamma_{\rm SDE}^{\rm SCD}$ for a GaAs-based quantum well are shown in Fig.~\ref{fig2}.
We assume only the first subband to be occupied, i.e., $n_\si(z)=N_s|\psi_{1\si}(z)|^2$,
and approximate the Kohn-Sham orbitals $\psi_{q,p\si}(z)$
entering Eq.~(\ref{gammaqw}) by the first two eigenstates of a harmonic oscillator with external potential
$\hbar^2 z^2/2m\lambda^4$. Furthermore, to lowest order in the non-adiabatic corrections  $\omega_\si$ can be
replaced with $\omega_{pq\si}$. For this system the parameters which govern the linewidth of the SDE mode are
$N_s$ and the quantum well curvature parameter $\lambda$. The latter determines both the excitation
frequency and the characteristic width of the ground-state density distribution.
The results in Fig. \ref{fig2} show that $\Gamma_{\rm SDE}^{\rm SCD}$ can be nonnegligible
(a large fraction of meV) for experimentally reasonable parameters, and $\Gamma_{\rm SDE}^{\rm SCD}/\hbar\omega$
can be of the order of few percents for a large range of curvature parameters and carrier densities.

For a specific $N_s$, the linewidth exhibits a well defined maximum as a function of $\lambda$.
The position of this maximum is determined by the competition of two distinct effects:
(i) The low-density saturation value of $\rho_{\uparrow \downarrow}$ {\em increases} with $\lambda$
[i.e. decreases with $\omega$, see Eq.~(\ref{rho_high_om})];
(ii) The average particle velocity {\em decreases} with $\lambda$ (i.e. decreases with the parabolic curvature).
The two effects give opposite contributions to the dissipation [see Eq.~(\ref{power_loss_gen})],
and the maximum occurs when the second effect takes over. Due to the density dependence of
$\rho_{\uparrow \downarrow}$ (see Fig.~\ref{fig1}),
a substantial contribution to the integrand in Eq.~(\ref{gammaqw}) can come from the lateral regions
of the quantum well, where the particle density is low. This is in contrast to the VE contribution,
which tends to be dominated by the high-density regions.

The above example shows that, even when other forms of damping, such as disorder and phonons, are drastically
reduced by careful selection of the system characteristics, the dissipation induced by SCD cannot
be avoided, due to its intrinsic nature. 

Eq.~(\ref{gammaqw}) suggests an experimental way to extract
the impact of SCD on spin dynamics by the optical measure of the linewidth of both charge- and spin-plasmons in
the same parabolic quantum well. Such measurements can be carried out using inelastic light scattering \cite{expe}.
Under the reasonable assumption that (i) extrinsic (ext) damping (non-magnetic impurities, phonons)
affect the CDE and SDE in the same way, and (ii) the viscoelastic term can be disregarded due to the parabolic system
geometry, we have
\be
\Gamma_{\rm SDE}-\Gamma_{\rm CDE}\approx  \left(\Gamma_{\rm SDE}^{\rm ext}+\Gamma_{\rm SDE}^{\rm SCD}\right)-
\left(\Gamma_{\rm CDE}^{\rm ext}\right)\approx \Gamma_{\rm SDE}^{\rm SCD} \:,
\ee
i.e., the SCD contribution to the spin-plasmon linewidth is given to a very good approximation by the difference of the
SDE and the CDE linewidths. This provides a valuable opportunity for comparison with microscopic models for the
transresistivity via Eq.~(\ref{gammaqw}), using the appropriate Kohn-Sham single-particle orbitals of the system.

In conclusion, we have presented a discussion of the power loss in a device due to dissipation of
spin-dependent currents induced by SCD forces. We have suggested a new, purely optical
method to measure the SCD effect in spin-density excitations in parabolic quantum wells.
In the $\omega \to 0$ limit, a particularly interesting application of our formalism would be
to describe the SCD intrinsic dissipation in spin-dependent transport through single molecular junctions \cite{Sai}.
As the broad effort in spintronics, quantum computation and transport in micro- and mesoscopic
systems continues, we expect a growing impact of the SCD effect in future applications.

\acknowledgments
This work was supported in part by DOE Grant DE-FG02-04ER46151, by the Petroleum Research Fund,
by the Nuffield Foundation Grant NAL/01070/G and
by the Research Fund 10024601 of the Department of Physics of the University of York.
C.~A.~U. is a Cottrell Scholar of Research Corporation.


\begin{thebibliography}{99}
\bibitem{note2}
As for the sign, we remind the reader that the Kohn-Sham {\it single} particle currents ${\bf j}_{pq\si}$
should not be confused with the velocities of the interacting system. In particular we underline that the former 
can have different
direction in respect to the the actual system current.
\end{thebibliography}
\end{document}